\documentclass[aps,showkeys,twocolumn,superscriptaddress,nofootinbib,preprint,10pt]{revtex4-2}
\usepackage{graphicx}
\usepackage{subfigure}

\graphicspath{{./fig/}}

\begin{document}
\title{High-field specific heat and entropy obtained from adiabatic temperature change}

\author{L. S. {Paix\~ao}}
\email{lpaixao@ifi.unicamp.br}
\affiliation{Instituto de F\'isica ``Gleb Wataghin'', Universidade Estadual de Campinas, 13083-859, Campinas, SP, Brazil.}

\author{E. O. Usuda}
\affiliation{Departamento de Ci\^encias Exatas e da Terra, Universidade Federal de S\~ao Paulo, 00972-270, Diadema, SP, Brazil.}

\author{W. Imamura}
\affiliation{Faculdade de Engenharia Mec\^anica, Universidade Estadual de Campinas, 13083-860, Campinas, SP, Brazil.}

\author{A. M. G. Carvalho}
\affiliation{Departamento de Engenharia Mec\^anica, Universidade Estadual de Maring\'a, 87020-900, Maring\'a, PR, Brazil.}
\affiliation{Departamento de Engenharia Qu\'imica, Universidade Federal de S\~ao Paulo, 09913-030, Diadema, SP, Brazil.}



\begin{abstract}
Specific heat and entropy are relevant thermodynamic properties, which may be used as macroscopic probes to microscopic properties of materials under ambient conditions and under high applied fields. However, the measurement of specific heat under intense external fields can be a challenging task, as well as to obtain the entropy in the same conditions. Here, we describe a method to obtain high-field specific heat and entropy from measurements of specific heat under ambient conditions and direct temperature change induced by adiabatic field changes. We derive straightforward thermodynamic equations to calculate the specific heat and entropy and our results agree satisfactorily with experimental data of specific heat under magnetic field, electric field, and pressure.
\end{abstract}

\maketitle

\section{Introduction}
Specific heat of solids is largely used to assess microscopic properties of materials. As external fields\footnote{Here we will consider pressure as a field of mechanical tension.} excite higher energy states of quasiparticles, such as phonons or magnons, measurable changes are observed on the material's specific heat, entropy, and other macroscopic quantities. As well known, with specific heat data from very low temperatures (near absolute zero) up to a temperature of interest, we can obtain the entropy in the same temperature range. Furthermore, if we measure specific heat under an external field, we can calculate the entropy under that field and the isothermal entropy change ($\Delta S_T$) between two states, at zero field and at applied field. Besides, the adiabatic temperature change ($\Delta T_S$) between two states, at zero field and at applied field, may also be calculated from entropy. $\Delta S_T$ and $\Delta T_S$ are the main parameters that characterize the caloric effect in materials. Generally, the higher the applied field, the higher $\Delta S_T$ and $\Delta T_S$.

Nevertheless, the measurement of specific heat under intense external fields can be a challenging task, as well to obtain the entropy in the same conditions. Specific heat under high magnetic fields is usually measured in commercial and in-house built calorimeters~\cite{prl-73-2744,acta.mater-107-1,adv.energy.mater-5-1901322}. However, it may represent a significant issue because of the magnetic forces on the samples, which result in undesired movements, leading to changes in thermal contact and time constants. Likewise, specific heat under high electric fields~\cite{prb-14-134,prb-16-433,epl-103-47001} and high mechanical tensions~\cite{j.low.temp.phys-120-245,jpcm-16-8905,thermochim.acta-446-66,nat.commun-6-8801} is often measured in in-house built calorimeters or through conformations of commercial calorimeters, and may be even more challenging.

Concerning the caloric materials, $\Delta S_T$ and $\Delta T_S$ may be obtained directly or indirectly. As mentioned before, $\Delta S_T$ and $\Delta T_S$ may be calculated from specific heat data and entropy, and it has been done for different caloric effects~\cite{j.appl.phys-86-565,prb-64-144406,rev.sci.instrum-81-104902,nat.commun-2-595,intermetallics-47-1,nat.commun-8-15715,pssb-255-1700422,prx-8-041035,jap-123-145101,nat.commun-10-1803}. Nevertheless, to the best of our knowledge, the reverse procedure has never been performed  systematically using $\Delta T_S$, i.e., a general method to obtain or estimate entropy and specific heat at applied fields from $\Delta T_S$ data has never been proposed besides a couple attempts to estimate specific heat under applied pressure~\cite{int.j.thermophys-6-101,j.appl.polym.sci-128-2269}.

Here, a method is described to obtain specific heat and entropy as a function of temperature under external fields, using zero-field specific heat and the adiabatic temperature change caused by the application or removal of the external field. The present method is general, since it is valid to any applied field, and relies entirely on straightforward thermodynamic equations. This method is validated using reported experimental data for different materials and caloric effects.

\section{High field specific heat}
An adiabatic process from an initial state described by temperature $T_i$ and field ${\cal F}_i$ (examples will be given where ${\cal F}$ is magnetic field, electric field, and pressure), to a final state described by temperature $T_f$ and field ${\cal F}_f$ is characterized by the condition
\begin{equation}
S(T_i,{\cal F}_i)=S(T_f,{\cal F}_f).
\end{equation}
Taking the partial derivative of the equation above, and recalling that $c_{\cal F}=T\partial S/\partial T$ is the specific heat under constant field, we arrive at
\begin{equation}
\frac{c_{\cal F}(T_i,{\cal F}_i)}{T_i}=\frac{c_{\cal F}(T_f,{\cal F}_f)}{T_f}\frac{\partial T_f}{\partial T_i}.
\label{eq_cf_geral}
\end{equation}
Mathematically, the final temperature $T_f=T_i+\Delta T_S(T_i,{\cal F})$ is a function of the initial temperature and the field. If $\Delta T_S=0$ then $\partial T_f/\partial T_i=1$. Therefore, from equation \ref{eq_cf_geral}, $c_{\cal F}(T_i,{\cal F}_i)=c_{\cal F}(T_i,{\cal F}_f)$. In other words, if an adiabatic change in the external field does not change the initial temperature of the material, the specific heat will not change as a function of the field at that temperature.

Without loss of generality, let us consider the particular case of a material that presents normal caloric effect, i.e., a material whose temperature rises when an external field is applied under adiabatic conditions and drops when the field is removed. Applying the field, the initial state is $T_i=T$ and ${\cal F}_i=0$, and the final state is $T_f=T+\Delta T_S$ and ${\cal F}_f={\cal F}$. Equation \ref{eq_cf_geral} reduces to
\begin{equation}
\frac{c_{\cal F}(T,0)}{T}=\frac{c_{\cal F}(T+\Delta T_S,{\cal F})}{T+\Delta T_S(T,{\cal F})}\left[1+\frac{\partial}{\partial T}\Delta T_S(T,{\cal F})\right].
\label{eq_cf_aplica_campo}
\end{equation}
On the other hand, while removing the field, the initial state is given by $T_i=T$ and ${\cal F}_i={\cal F}$, and the final state is $T_f=T-|\Delta T_S|$ (the temperature change is negative) and ${\cal F}_f=0$. Equation \ref{eq_cf_geral} then becomes
\begin{equation}
\frac{c_{\cal F}(T,{\cal F})}{T}=\frac{c_{\cal F}(T-|\Delta T_S|,0)}{T-|\Delta T_S(T,{\cal F})|}\left[1-\frac{\partial}{\partial T}|\Delta T_S(T,{\cal F})|\right].
\label{eq_cf_remove_campo}
\end{equation}
Equations \ref{eq_cf_aplica_campo} and \ref{eq_cf_remove_campo} allow us to obtain the specific heat under an external field for materials that we have available zero-field specific heat and adiabatic temperature change data as a function of temperature.

From the total differential of entropy it is possible to write the well known equation
\begin{equation}
\Delta T=-\int_{{\cal F}_i}^{{\cal F}_f}\frac{T}{c_{\cal F}}\left.\frac{\partial S}{\partial{\cal F}}\right|_Td{\cal F},
\end{equation}
where, via a Maxwell relation, $\partial S/\partial{\cal F}$ is usually written in terms of the quantity that forms a conjugate pair with the field $\cal F$. This equation is largely used to indirectly estimate the temperature change from the knowledge of the specific heat. What we propose here is the converse method, to evaluate the specific heat under applied field from the directly measured temperature change.

It is worth pointing out that integrating the specific heat to obtain the entropy curves under zero field and under applied field, evaluating the expression
\begin{equation}
S(T,{\cal F})=S_{ref}({\cal F})+\int_{T_{ref}}^T\frac{c_{\cal F}(T',{\cal F})}{T'}dT',
\label{eq_entropia}
\end{equation}
where $S_{ref}({\cal F}>0)$ may be apropriately chosen in order to satisfy the condition $S(T,0)=S(T+\Delta T_S,{\cal F})$ at high temperautre.

\section{Applying the method to experimental data}
\subsection{Gadolinium}
\begin{figure}[htb]
\centering
\subfigure{\includegraphics[width=.45\textwidth]{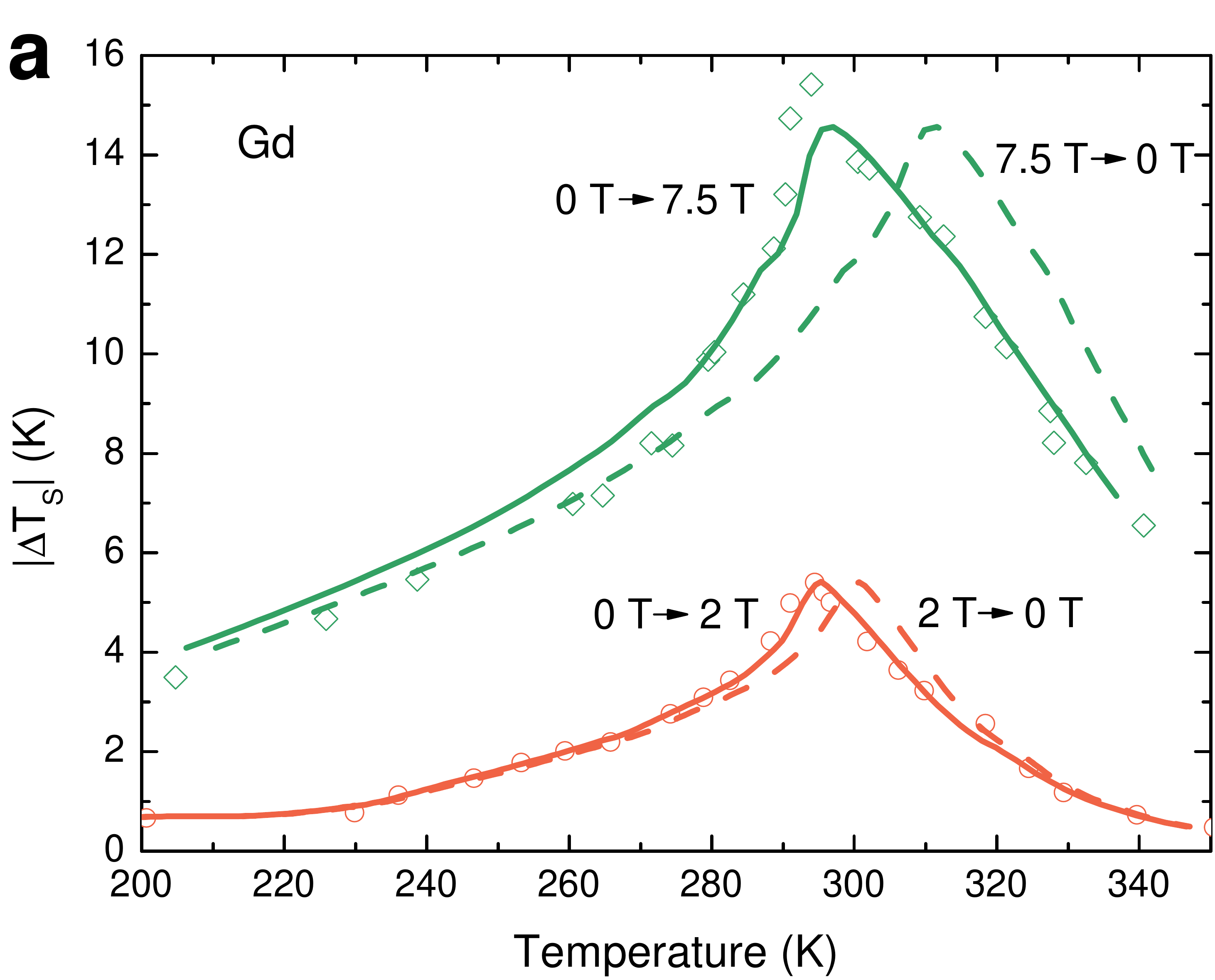}}\\
\subfigure{\includegraphics[width=.45\textwidth]{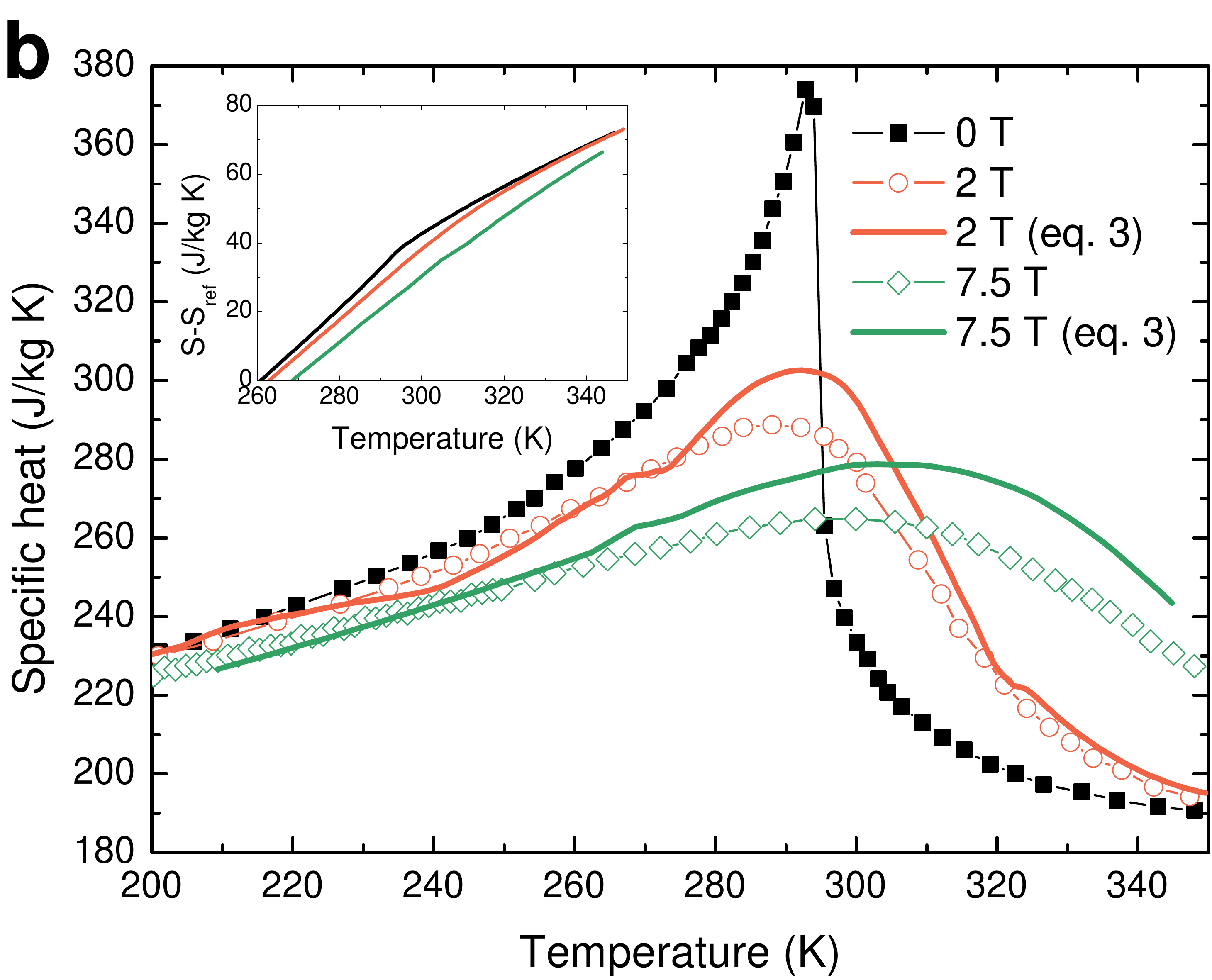}}
\caption{(a) Direct temperature change of Gd reproduced from reference \onlinecite{prb-57-3478} (symbols) and temperature change evaluated from the entropy curves (lines). (b) Specific heat data reproduced from reference \onlinecite{prb-57-3478} (symbols) and specific heat evaluated with equation \ref{eq_cf_aplica_campo} (lines). Inset: entropy-temperature diagram built from the calculated specific heat.}
\label{fig1}
\end{figure}

There is sufficient data on literature to test the equations above. Let us take as a first example gadolinium, which is a magnetic material that presents a second order transition between ferromagnetic and paramagnetic phases at 293 K. At the transition, a temperature change of almost 16 K~\cite{prb-57-3478} upon the adiabatic application of a magnetic field of 7.5 T is observed. \figurename\ \ref{fig1}a presents the adiabatic temperature change signature on a ferro/paramagnetic transition measured on a polycrystalline sample~\cite{prb-57-3478}. The zero-field specific heat of single-crystal Gd~\cite{prb-57-3478}, on \figurename\ \ref{fig1}b, presents the transition at the same temperature. Applying equation \ref{eq_cf_aplica_campo} to numerically evaluate the specific heat under constant magnetic field we see a reasonable agreement with the experimental data.

Once the specific heat at applied magnetic field is available, it is possible to build the entropy-temperature diagram shown on the inset of \figurename\ \ref{fig1}b using equation \ref{eq_entropia}. From the entropy, we obtained the temperature change upon adiabatic application and removal of the magnetic field (also shown on \figurename\ \ref{fig1}a). The experimental and calculated $\Delta T_S$ are in a good agreement.

\subsection{Manganese arsenide}
Let us focus on another well known magnetocaloric material. MnAs presents large temperature changes, as reported by L. Tocado \emph{et al.}~\cite{j.therm.anal.calorim-84-213}, due to its first order magneto-structural transition, between a paramagnetic orthorhombic phase and a ferromagnetic hexagonal phase. Specific heat data reported by the same authors show that the transition, at zero field, occurs at 315.7 K on heating and at 305.7 K on cooling.

\begin{figure}[htb]
\centering
\subfigure{\includegraphics[width=.45\textwidth]{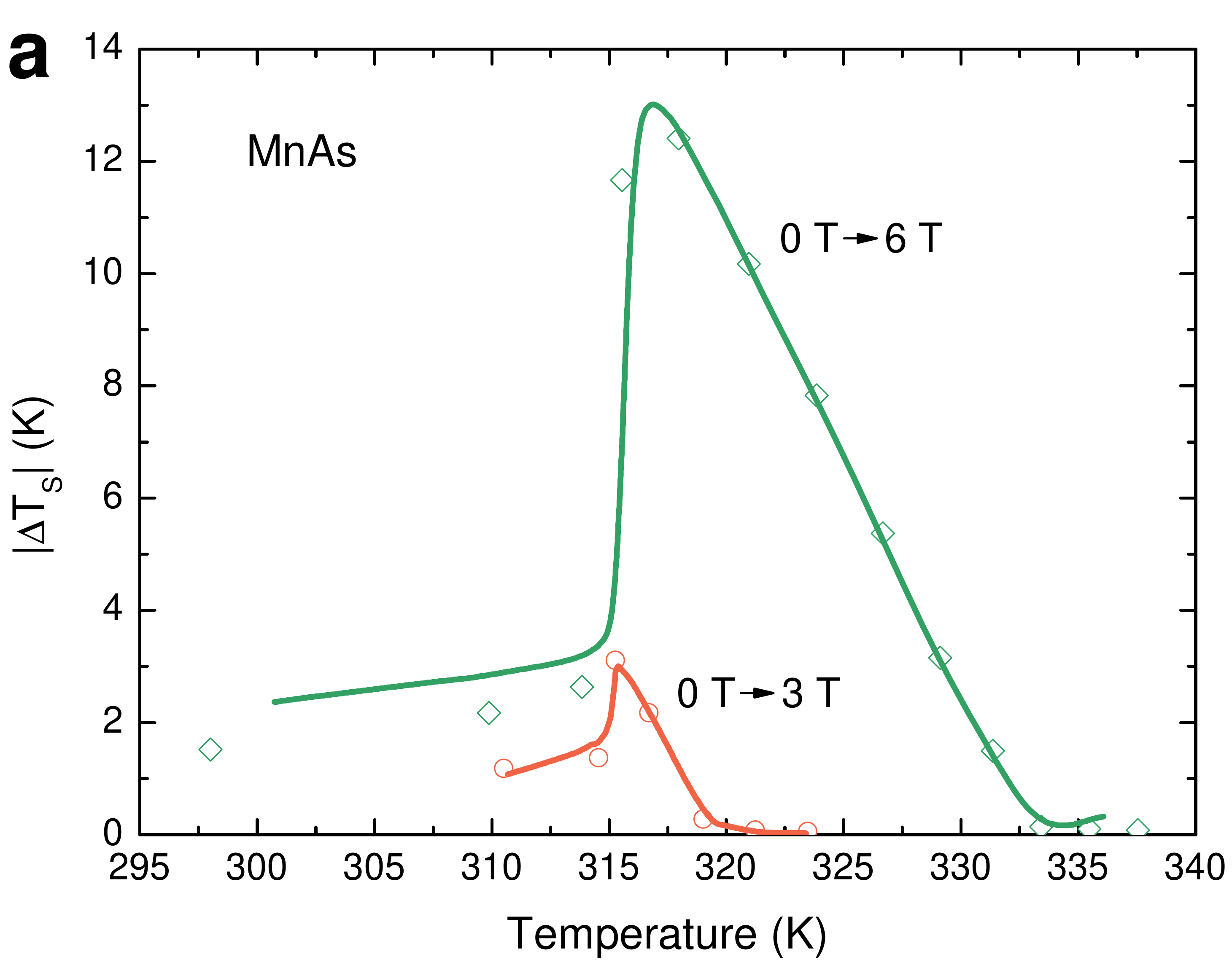}}\\
\subfigure{\includegraphics[width=.45\textwidth]{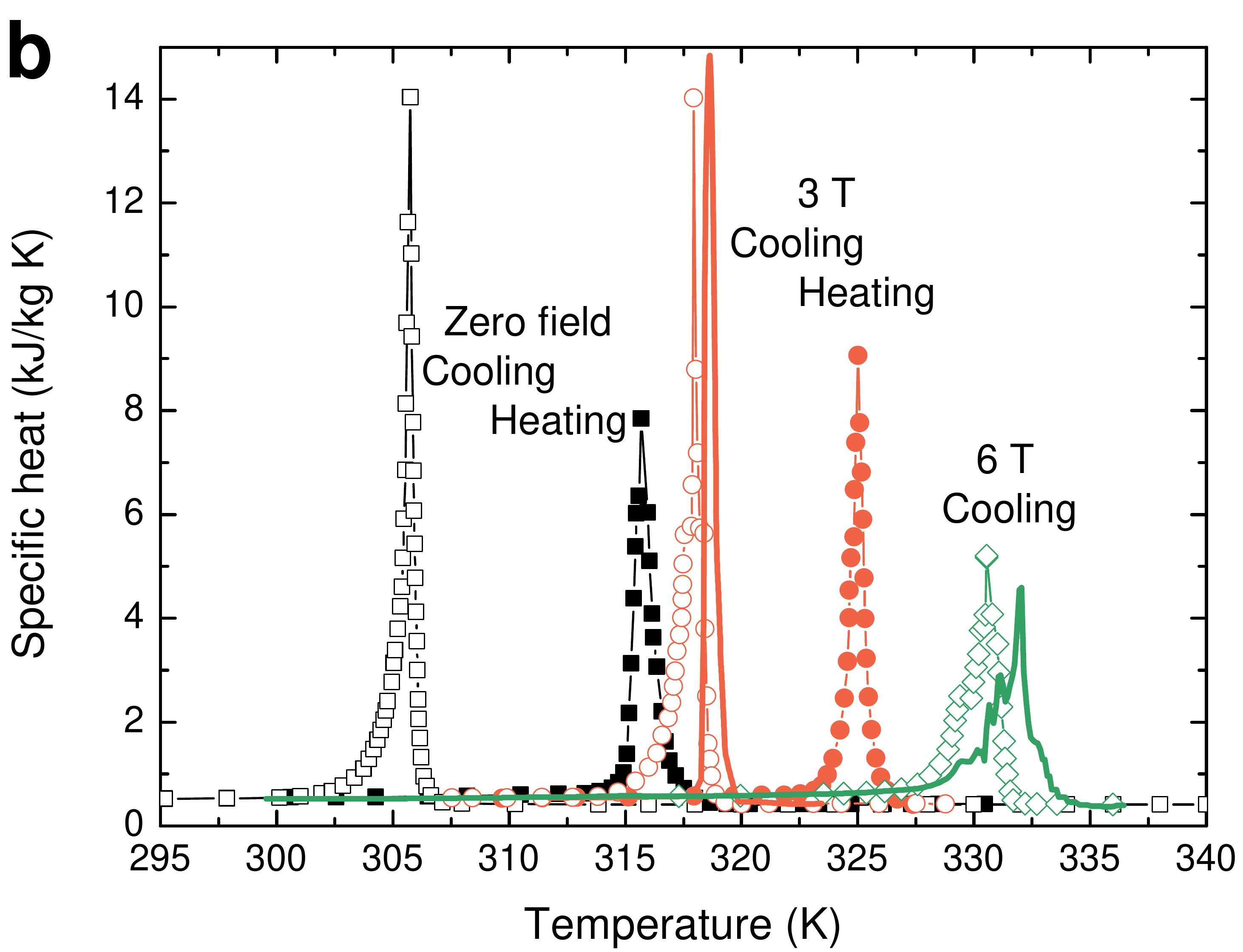}}
\caption{(a) Direct temperature change of MnAs reproduced from reference \onlinecite{j.therm.anal.calorim-84-213} (symbols) and temperature change evaluated from the entropy curves (lines). (b) Specific heat data reproduced from reference \onlinecite{j.therm.anal.calorim-84-213} (symbols) and specific heat evaluated with equation \ref{eq_cf_aplica_campo} (lines).}
\label{fig2}
\end{figure}

The adiabatic temperature change shown on \figurename\ \ref{fig2}a (reproduced from reference \onlinecite{j.therm.anal.calorim-84-213}) was measured during application of the magnetic field. Also on \figurename\ \ref{fig2}a, the temperature change calculated from the entropy curves, as described above (equation \ref{eq_entropia}). Only the curves for the application of the field are shown for a simple comparison with the experimental data available.

Therefore, applying equation \ref{eq_cf_aplica_campo} to the zero-field specific heat we obtain the specific heat under 3 T and 6 T, shown on \figurename\ \ref{fig2}b. We see that the calculated high-field specific heat matches the experimental cooling curve, when calculated from the zero-field heating specific heat curve. In other words, the experimental temperature change is the temperature difference between the high-field cooling entropy curve and the zero-field heating curve. T. Christiaanse \emph{et al.}~\cite{j.phys.d-50-365001} performed a detailed study of the first order transition of a Mn-Fe-P-Si magnetic alloy and reached the same conclusion. They also provide other references, like \onlinecite{jap-113-173510} and \onlinecite{jap-116-063903}, that report the same phenomenon on related materials.

\subsection{Barium titanate}
\begin{figure}[htb]
\centering
\subfigure{\includegraphics[width=.5\textwidth]{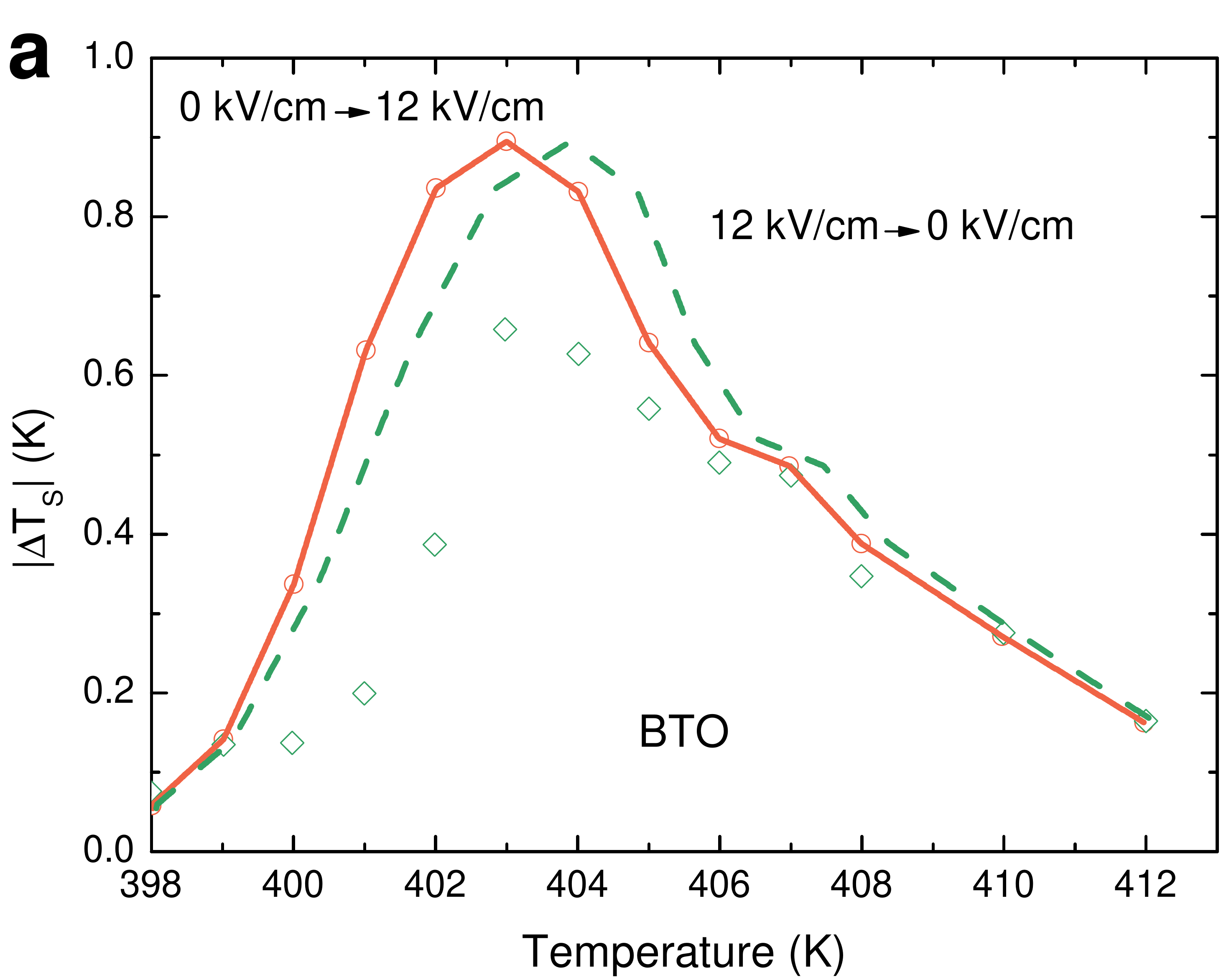}}\\
\subfigure{\includegraphics[width=.5\textwidth]{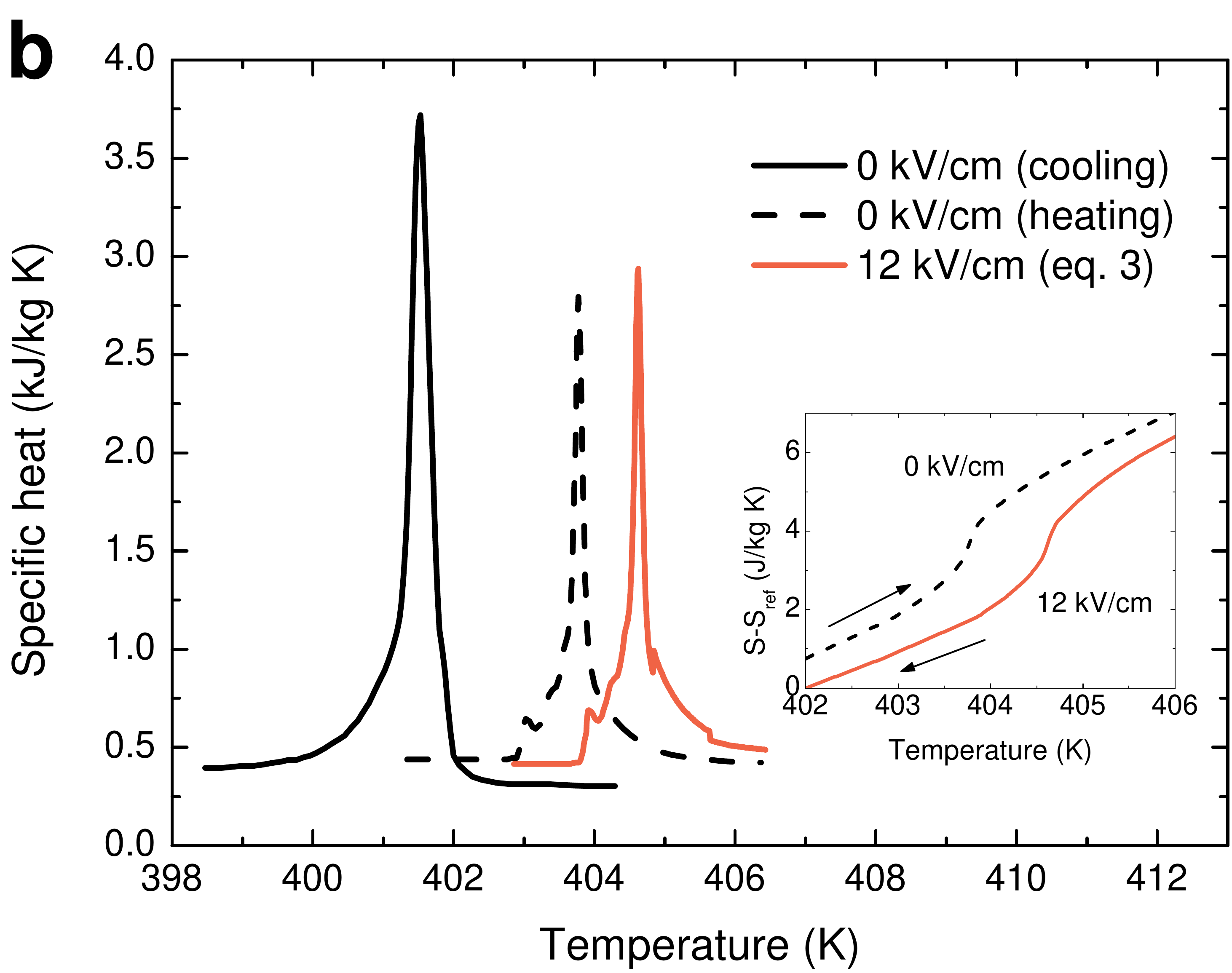}}
\caption{(a) Direct temperature change of BTO, reproduced from reference \onlinecite{adv.mater-25-1360} (symbols) and temperature change evaluated from the entropy curves (lines). (b) Specific heat under zero field reproduced from reference \onlinecite{adv.mater-25-1360}, and specific heat under electric field evaluated from equation \ref{eq_cf_aplica_campo}.}
\label{fig3}
\end{figure}

Let us now investigate the specific heat of samples under external electric fields. X. Moya \emph{et al.} reported the electrocaloric effect of barium titanate BaTiO$_3$ (BTO) single crystals~\cite{adv.mater-25-1360}. BTO undergoes a first order ferro/paraelectric transition around 400 K. The adiabatic temperature changes due to application and removal of the electric field, shown on \figurename\ \ref{fig3}a, present considerable discrepancy; not only the maximum $\Delta T_S$ is displaced, but the maximum $\Delta T_S$ value is lower during field removal. Such difference is attributed to the thermal hysteresis of the transition, since the removal of the field does not fully drive the transition to the paraelectric phase in the hysteresis region. The calculated temperature change induced by removal of the field peaks at a higher temperature compared to the temperature change induced by the application of the field, following the experimental trend.

The specific heat of BTO under an electric field of 12 kV/cm was evaluated from equation \ref{eq_cf_aplica_campo} using the zero-field heating specific heat. According to the previous discussion (regarding MnAs), the calculated specific heat of BTO, shown on \figurename\ \ref{fig3}b, must correspond to the cooling process. The entropy curves shown on the inset of \figurename\ \ref{fig3}b represent the fully paraelectric and the fully ferroelectric phases, which explains the calculated $\Delta T_S$ induced by the field removal being larger than the experimental counterpart.

\subsection{Vulcanized natural rubber}
\begin{figure}[htb]
\centering
\subfigure{\includegraphics[width=.5\textwidth]{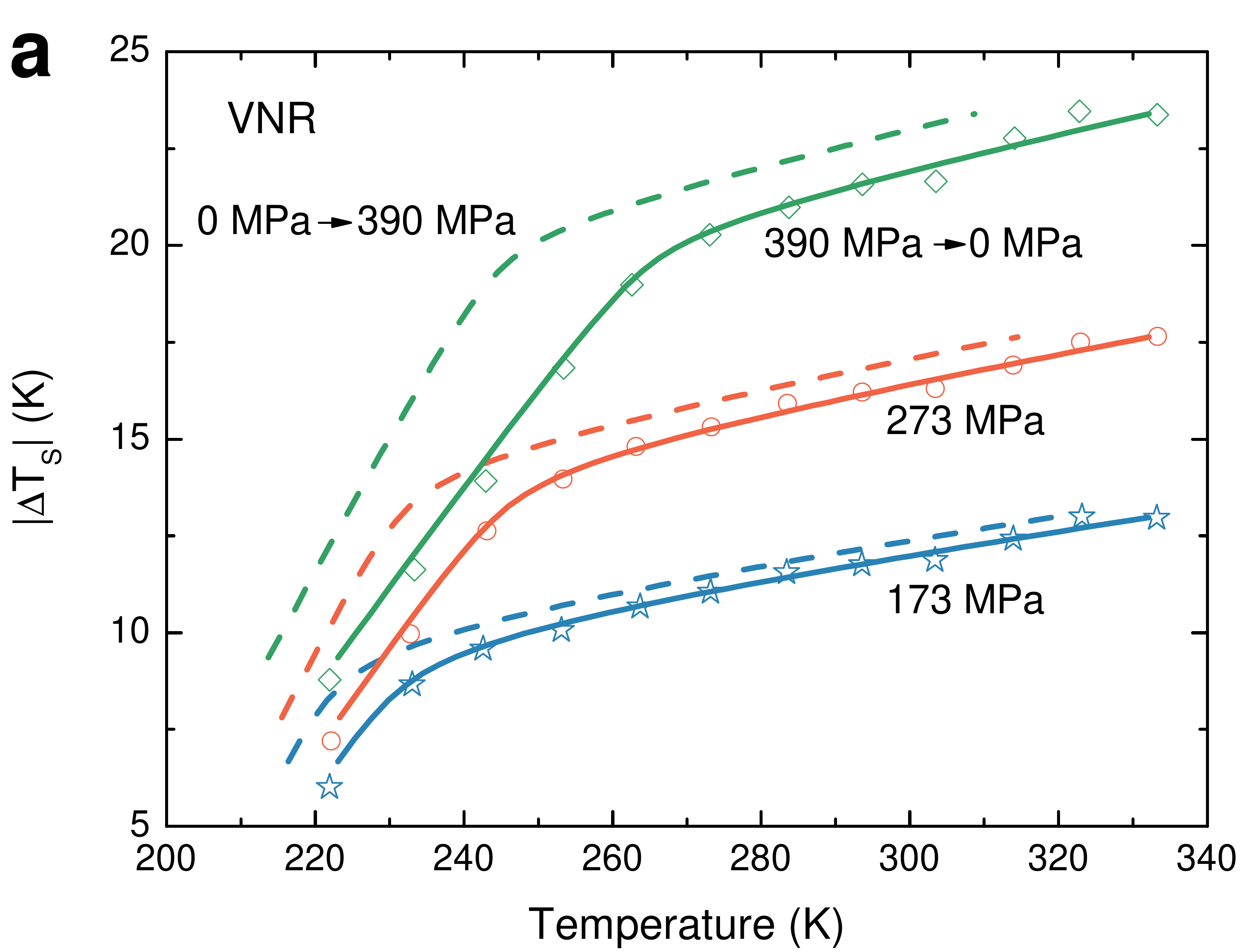}}\\
\subfigure{\includegraphics[width=.5\textwidth]{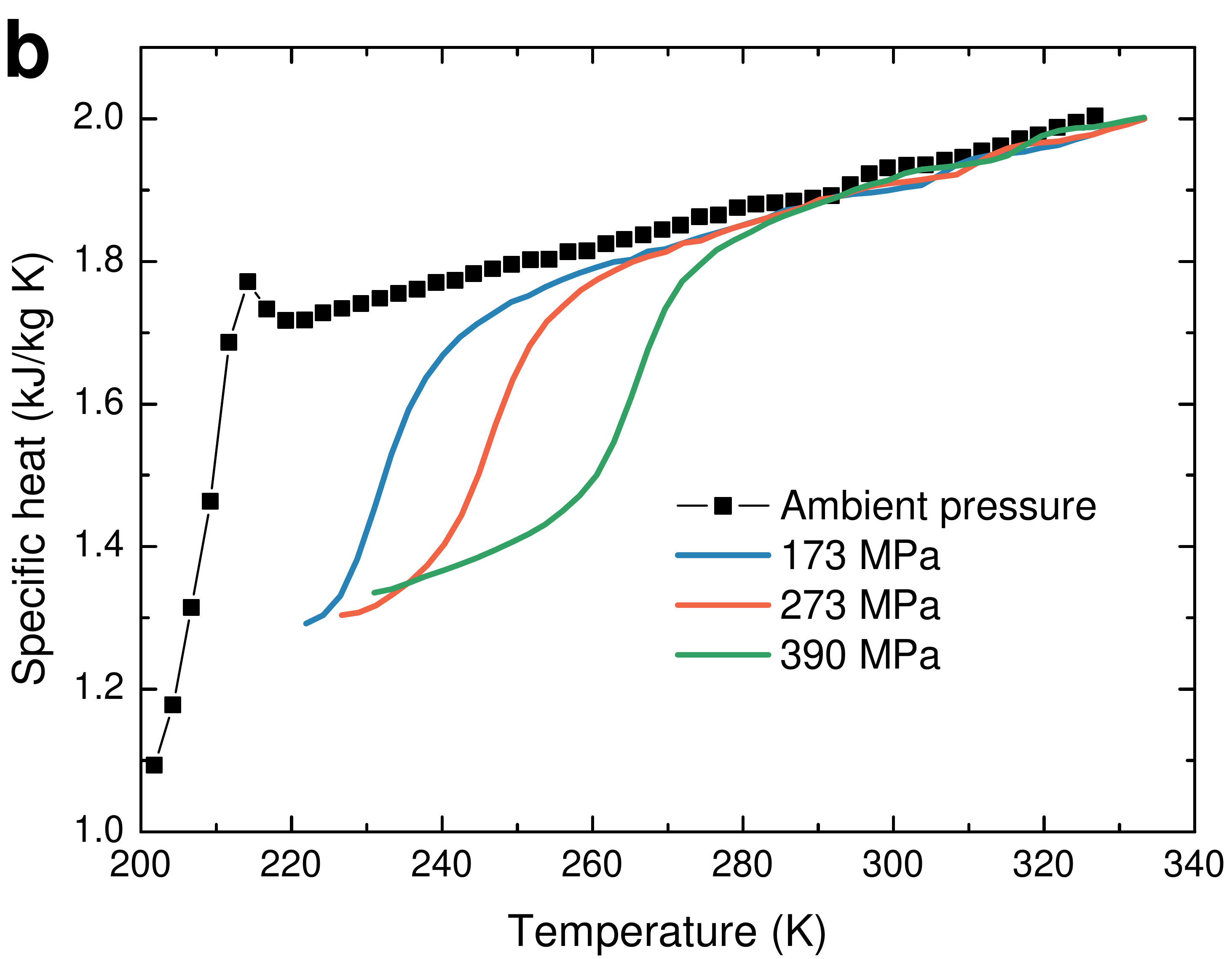}}
\caption{(a) Direct temperature change of VNR, measured during adiabatic decompression (symbols), and calculated from the entropy curves (lines). (b) Ambient pressure specific heat (symbols) measured under a 10 K/min heating rate; high-pressure specific heat (lines) calculated with equation \ref{eq_cf_remove_campo}.}
\label{fig4}
\end{figure}

Vulcanized natural rubber (VNR) is an elastomeric polymer, whose barocaloric properties have been thoroughly investigated recently~\cite{rev.sci.instrum-88-046103,macro.lett-7-31}. Unlike other materials that manifest caloric effects near phase transitions, VNR presents considerable temperature changes over a wide temperature range. As other polymers, VNR exhibits a glass transition ($T_g$) at 210 K at ambient pressure, and the glass transition linearly increases to 266 K at 390 MPa. Below $T_g$, the temperature change drops quickly, due to the reduced mobility of the polymer chains in the glassy state. \figurename\ \ref{fig4}a presents the temperature change of a new sample of VNR measured according to the procedure described in reference \onlinecite{macro.lett-7-31}. Ambient pressure specific heat, shown on \figurename\ \ref{fig4}b, was measured on a differential scanning calorimeter model DSC 214 Polyma, from Netzsch. The measurement was performed under a 10 K/min heating rate.

In order to avoid numerical fluctuations related to measurement resolution a fitting was applied to the experimental temperature change, according to the equation described in the Supporting Information. Solid and dashed lines on \figurename\ \ref{fig4}a are the calculated temperature change on decompression and compression, respectively, obtained from the entropy curves. The calculated values are consistently higher during compression, in accordance with the behavior observed up to 87 MPa~\cite{rev.sci.instrum-88-046103}. High-pressure specific heat, shown in \figurename\ \ref{fig4}b, was evaluated from equation \ref{eq_cf_remove_campo} using the fitted decompression $\Delta T_S$ (shown in the Supporting Information).

\section{Conclusion}
In summary, the current paper presents a method to obtain the specific heat and entropy of materials under applied fields. Such specific heat and entropy are evaluated from the specific heat under ambient conditions and the temperature change under adiabatic application or removal of the field. The equations derived were applied to experimental data from different materials under the influence of magnetic field, electric field or pressure. Besides, the materials experienced magnetic, magnetostructural, electric, and glass transitions. In all situations, the calculated specific heat agrees with the experimental data available. Furthermore, also entropy and isothermal entropy change agree satisfactorily. Under a first-order transition, specific heat and entropy present thermal hysteresis, and it was observed that the experimental temperature change corresponds to the temperature difference along an isentropic line, between the zero-field heating entropy curve and the high-field cooling entropy curve, in accordance with previous reports. From the zero-field and high-field entropy curves (integrated from the corresponding specific heat), it is possible to evaluate the adiabatic temperature change and the isothermal entropy change corresponding to the application and removal of the respective field. The calculated temperature change must match the experimental one, and is a good measure of the accuracy of the method. Moreover, the present method may be used to check the quality and consistency of experimental data sets, in this case, experimental specific heat, entropy, adiabatic temperature change and isothermal entropy change.

\begin{acknowledgments}
This study was financed in part by the Coordena\c{c}\~ao de Aperfei\c{c}oamento de Pessoal de N\'ivel Superior (CAPES) -- Finance code 001, the Conselho Nacional de Desenvolvimento Científico e Tecnológico (CNPq), and the Funda\c{c}\~ao de Amparo \`a Pesquisa do Estado de S\~ao Paulo (FAPESP).
\end{acknowledgments}

\appendix
\section*{Supporting information}
\subsection*{Entropy change}
\begin{figure}[htb]
\centering
\subfigure{\includegraphics[width=.5\textwidth]{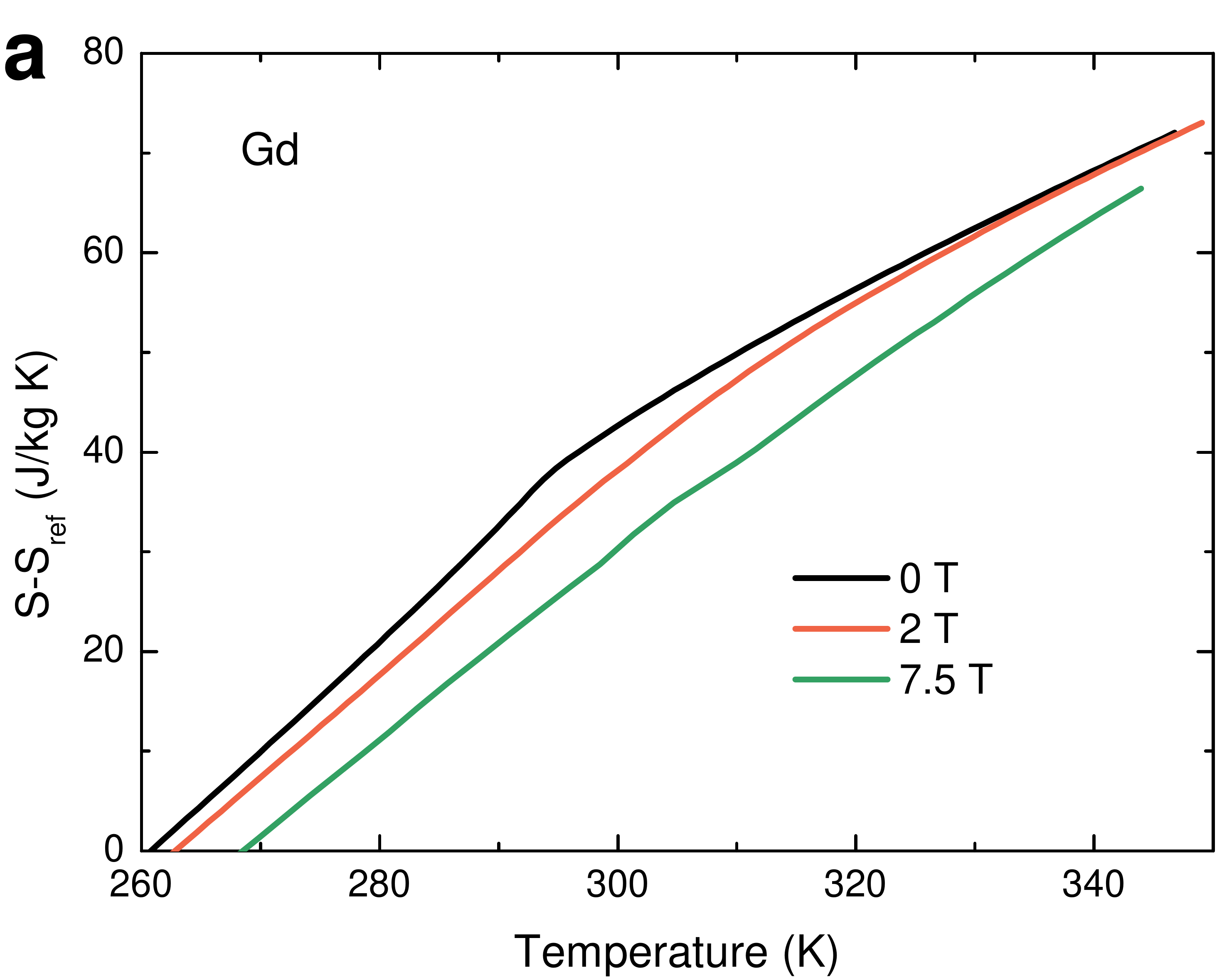}}\hfill
\subfigure{\includegraphics[width=.5\textwidth]{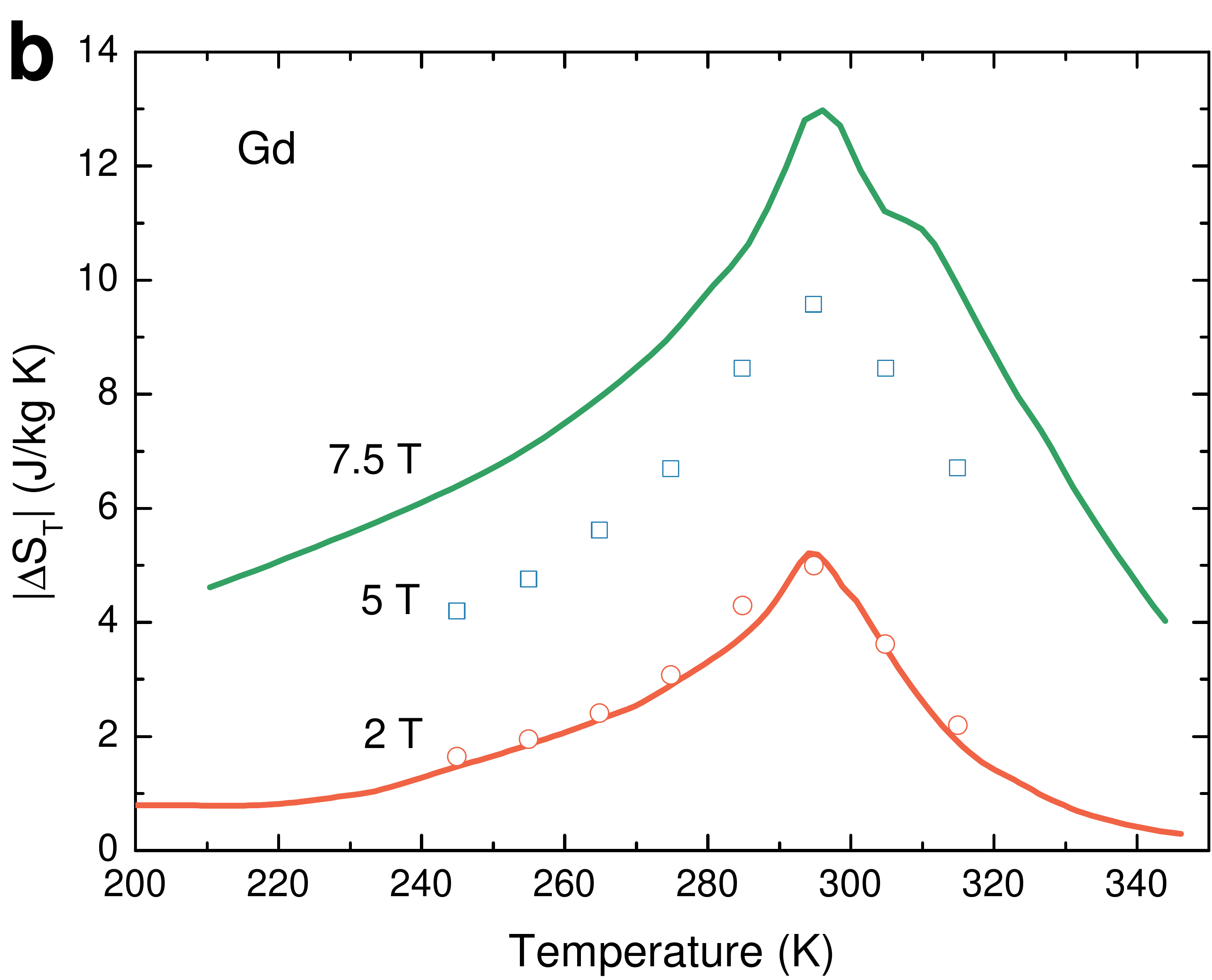}}
\caption{(a) Entropy of Gd around the second order transition temperature. (b) Isothermal entropy change: lines were calculated from the entropy shown on (a), and symbols (reproduced from reference \cite{prb-57-3478}) were obtained from magnetization measurements.}
\label{figs1}
\end{figure}

Entropy-temperature diagrams, like the one shown on \figurename\ \ref{figs1}a below for Gd (which also appears on Fig. 1b of main text), can be used to obtain the isothermal entropy change as a function of temperature. Such quantity is usually evaluated indirectly from magnetization measurements (or the quantity conjugated to the corresponding applied field) via Maxwell relation. \figurename\ \ref{figs1}b presents the comparison between the entropy change of Gd calculated from the entropy curves and the entropy change obtained via Maxwell relation~\cite{prb-57-3478}. Under a 2 T magnetic field change, the calculated and the indirect entropy changes of Gd present remarkable agreement.

\begin{figure}[htb]
\centering
\subfigure{\includegraphics[width=.5\textwidth]{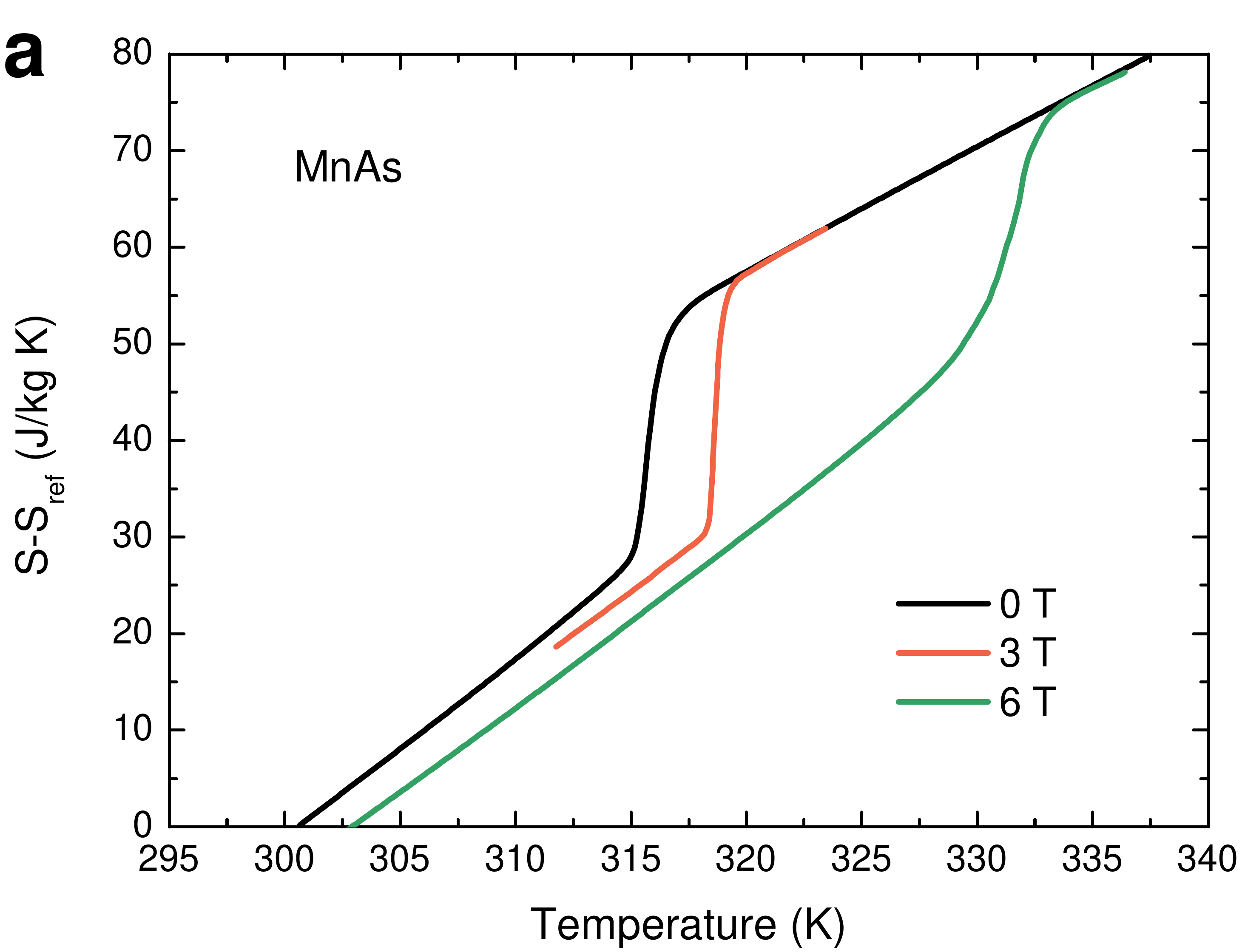}}\hfill
\subfigure{\includegraphics[width=.5\textwidth]{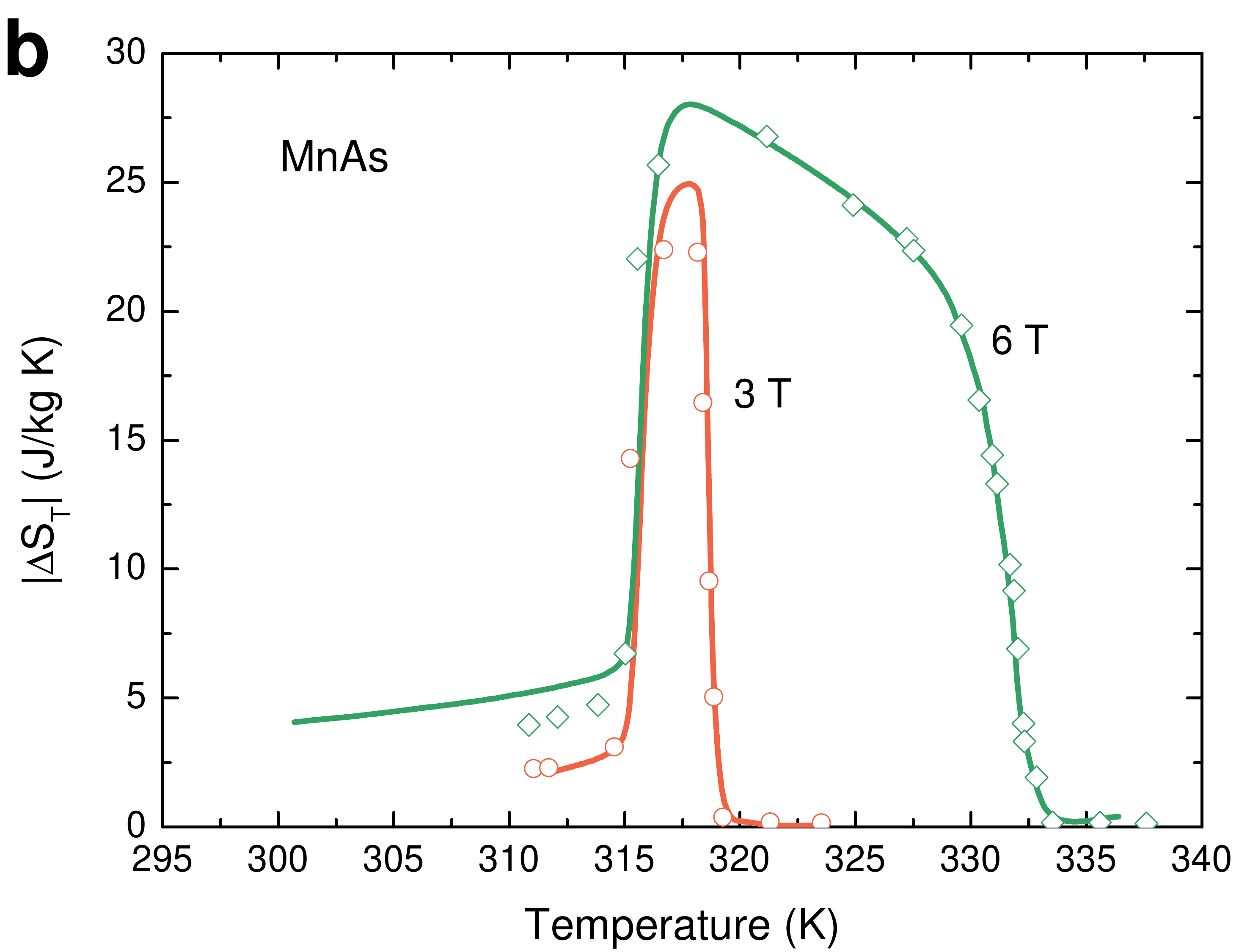}}
\caption{(a) Entropy of MnAs as a function of temperature. (b) Isothermal entropy change: lines were calculated from the entropy curves, and symbols (reproduced from reference \cite{j.therm.anal.calorim-84-213}) were obtained from specific heat and direct temperature change.}
\label{figs2}
\end{figure}

It is also possible to construct the entropy-temperature diagram for a first order transition, as shown on \figurename\ \ref{figs2}a for MnAs. Again, the entropy curves were obtained integrating the specific heat, and the high-field entropy curves were placed following the experimental temperature change (the method is described in the main text). On \figurename\ \ref{figs2}b, the entropy change evaluated from the entropy curves agrees noticeably with the values from literature \cite{j.therm.anal.calorim-84-213}.

As discussed in the main text, the method reported here is reliable for obtaining the high-field specific heat of materials. Besides, as we now show, the entropy change evaluated from the entropy-temperature diagrams is consistent with previous methods, irrespective of the order of the transitions involved.

\subsection*{Temperature change of VNR around $T_g$}
Temperature change of VNR around the glass transition ($T_g$) is described by the empirical expression
\begin{equation}
\left(\frac{\Delta T_S(t)-\Delta T_S(0)}{\tau}-\beta t\right)\frac{\pi}{\alpha}=\frac{1}{2}\ln(1+t^2)-t\arctan(t),
\label{eq_fit_deltat}
\end{equation}
with
\begin{equation}
t=\frac{T-T_g}{\tau}.
\end{equation}
Above, $\alpha$ is the change in slope during the transition, $\beta$ is the slope at $T_g$, and $\tau$ is the width of the transition. Note that $t$, $\alpha$, and $\beta$ are dimensionless, while $\tau$ has the dimension of temperature.

\begin{figure}[htb]
\centering
\includegraphics[width=.5\textwidth]{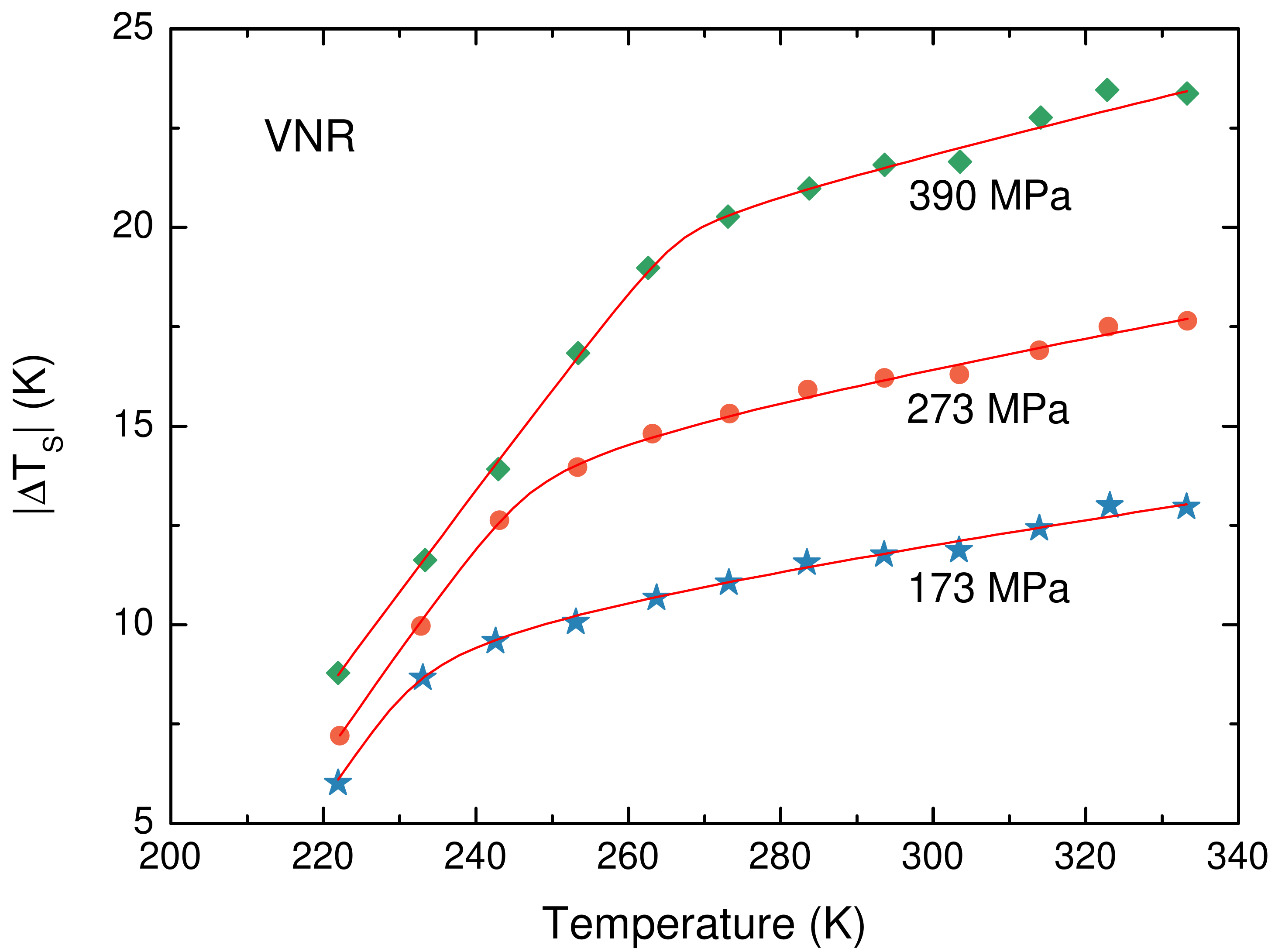}
\caption{Temperature change of VNR at decompression and the corresponding fittings of equation \ref{eq_fit_deltat}.}
\label{figs3}
\end{figure}

\figurename\ \ref{figs3} presents the experimental temperature change and the corresponding fitting, according to equation \ref{eq_fit_deltat}, for pressures 173 MPa, 273 MPa, and 390 MPa. The fitting parameters are shown in \tablename\ \ref{tabs1}. The fitted $\Delta T_S$ were used in the calculation of the specific heat discussed in the main text.

\begin{table}[htb]
\centering
\caption{Fitting parameters for the empirical temperature change expression.}
\begin{tabular}{ccccc}
\hline\hline

Pressure (MPa) & $\alpha$ & $\beta$ & $\tau$ (K) & $T_g$ (K)\\

\hline

173 & 0.26(2) & 0.16(1) & 4(1) & 234(1)\\
273 & 0.29(1) & 0.170(5) & 7(1) & 246(1)\\
390 & 0.22(2) & 0.159(4) & 4(3) & 265(1)\\

\hline\hline
\end{tabular}
\label{tabs1}
\end{table}


\end{document}